\documentstyle[aps,preprint]{revtex}
\begin{document}
\draft

\preprint{KAIST-TH-97/6}

\title{$S$-wave sector of type IIB supergravity
       on $S^1 \times T^4$ }
\author{Youngjai Kiem$^{(a)}$, Chang-Yeong Lee$^{(b)}$ 
        and Dahl Park$^{(a)}$ }
\address{ $^{(a)}$ Department of Physics \\
                   KAIST\\
                   Taejon 305-701, Korea \\
          $^{(b)}$ Department of Physics \\
                  Sejong University\\
                  Seoul 143-747, Korea \\
         }
\maketitle

\begin{abstract}
We consider the type IIB supergravity in ten dimensions 
compactified on $S^1 \times T^4$, with intersecting one and 
five D-branes in the compact dimensions.  By imposing the 
spherical symmetry in the resulting five dimensional theory, 
we further reduce the $s$-wave sector of the theory to a 
two dimensional dilaton gravity.  Via this construction, the 
techniques developed for the general two dimensional dilaton 
gravities are applicable in this context.  Specifically, we 
obtain the bosonic sector general static solutions.  In addition 
to the well-known asymptotically flat black hole solutions, 
they include solutions with naked singularities and 
non-asymptotically flat black holes.
\end{abstract}


\section{Introduction}

One of the most exciting recent developments in black 
hole physics is the microscopic counting of the 
black hole entropy from string theory \cite{entropy}.
For a class of extremal and near extremal black holes 
in supergravity theories, the area law for the entropy 
and the quantum emission rate from near extremal black 
holes were derived from the microscopic considerations 
of string theory \cite{malda}.  The simplest and
illuminating example comes from the type IIB 
supergravity compactified on $S^1 \times T^4$ 
with intersecting one and five $D$-branes \cite{callan}.  
The classical supergravity solutions 
in this case can be related to the microscopic system 
involving $D$-branes.  The electric and 
magnetic charges of the RR three
form field strength can be interpreted as the wrapping 
number of $D$ one-branes around $S^1$ and the wrapping
number of $D$ five-branes around $S^1 \times T^4$,
respectively.  The Kaluza-Klein charge along the 
circle $S^1$ corresponds to the massless excitations 
of $D$-branes represented by the open strings connecting 
the one and five $D$-branes.  

     Throughout these lines of investigation, it is of 
prime importance to obtain the classical black hole
solutions of the supergravity theories, and considerable 
efforts were devoted to this issue \cite{cvetic2} \cite{hs}.  
In Ref. \cite{hms}, a six-parameter class of black
hole solutions in the type IIB supergravity compactified
on $S^1 \times T^4$ were constructed; the parameters
represent the black hole mass, one-brane 
charge, five-brane charge, Kaluza-Klein momentum, the
asymptotic circle size and the asymptotic four-torus
size.  One important tool in the 
construction of the supergravity solutions is the 
solution generating technique; starting
from a particular generating solution, the general 
black holes solutions with {\em no-hair} property are
obtained by maximal $S$- and $T$-duality 
transformations \cite{cvetic2} \cite{cvetic1} 
\cite{horowitz}. 

Considering these developments, it is of some interest
to find the general static solutions compatible 
with the intersecting one-brane and five-brane 
configurations on $S^1 \times T^4$.  First,
that kind of analysis will show in detail how 
potential hairs are prevented for black hole solutions.
Secondly, if there exist self-dual solutions under the
solution generating transformations mentioned above, 
they can not be obtained by the solution generating 
techniques.  In this paper, by restricting our 
attention only to the $s$-wave sector of the 
five dimensional theory (obtained by compactifying
type IIB theory on $S^1 \times T^4$), we achieve
this task.

In Section II, we observe that the $s$-wave
sector of the type IIB supergravity on 
$S^1 \times T^4$ becomes a particular type of 
two dimensional dilaton gravity theory \cite{dilaton}.
Due to the vastly simplified dynamics in two
dimensions, some analytic treatments are possible
\cite{dil2}.  For example, the procedures for 
getting the general static solutions for a large class 
of two dimensional dilaton gravity
theories were given in Ref. \cite{kp}.
Following that method, in Section III, we 
obtain the general static solutions labeled by
fifteen parameters without any assumptions about 
the global structure of the space-time.
The solution space has very rich structures
composed of many sectors.  We present
a detailed study in a particular 
sector that, with an appropriate restriction
of parameters, includes the solutions of
Ref. \cite{hms}.  We find that unless the 
charges of modulus fields are set to
certain values (determined by gauge charges
and the asymptotic values of the modulus fields)
we have naked singularities.  Thus, if we make 
a restriction on the parameters to avoid naked 
singularities, we recover the asymptotically flat 
black hole solutions of Ref. \cite{hms}.
In addition, our general solutions in that sector
also include non-asymptotically flat black hole 
type solutions.  A similar result is already 
known in the context of the four dimensional
dilaton gravity; in that theory, there are 
asymptotically flat black hole solutions by 
Garfinkle, Horowitz and Strominger
\cite{ghs} and non-asymptotically flat black
hole type solutions
of Ref. \cite{chm}.  In Section IV, we
comment on some more aspects of the general
static solutions and our approach.

\section{Dimensional reduction $10 \rightarrow 6 \rightarrow
         5 \rightarrow 2$ }
We start with the bosonic sector of the ten-dimensional
type IIB supergravity compactified on $S^1 \times T^4$.
\begin{equation}
I^{(10)} = \int d^{10} x \sqrt{-g^{(10)}} ( e^{-2 \phi}
R^{(10)} + 4 e^{-2 \phi} (D \phi )^2
- \frac{1}{12} H^{(10) 2 } )
\end{equation}
We include in our action the metric $g^{(10)}_{\mu \nu}$,
the dilaton $\phi$ and the R-R three-form field strength
$H^{(10)}$.  Among the bosonic fields of the type IIB
theory, we set the NS-NS three-form field strength, 
the R-R one-form field strength and the R-R self-dual
five-form field strength zero, since we consider the 
configuration of intersecting one and five D-branes
in this paper.  The signature for the metric and
other conventions are explained in the Appendix A.

We first compactify the ten dimensional theory on 
the four-torus $T^4$.  We assume the maximal 
symmetry on the torus (the Euclidean
Poincare symmetry for the covering space $R^4$
and the hypercubic lattice for the discrete
action on $R^4$)  
and, therefore, we include only single modulus
$\psi$ for the $T^4$.  The ten dimensional
metric is thus given in the block diagonal form 
\begin{equation}
ds^2 = g^{(6)}_{\alpha \beta} dx^{\alpha} 
dx^{\beta} + e^{\psi} dx^m dx^m
\end{equation}
where the Greek indices $\alpha$ and $\beta$ 
run from zero to five and the Latin index $m$
runs from six to nine with the flat Euclidean metric.  
The scalar field 
$\psi$ measures the size of the four-torus.  
The gauge invariant three 
form field strength $H^{(10)}$ should also respect 
the assumed symmetry.  Consequently, we set
$H^{(10)}_{\alpha \beta m} = 0$, 
$H^{(10)}_{\alpha m n } = 0$ and 
$H^{(10)}_{m n l} = 0$ and retain only 
$H^{(10)}_{\alpha \beta \gamma} = 
H^{(6)}_{\alpha \beta \gamma}$, which is both a 
scalar on the $T^4$ and a three form field strength on
the six-dimensional transversal manifold.  Under 
the assumed symmetry, we note that the invariant 
objects are the scalars that do not depend on
$x^m$ coordinates and the volume 
form $\epsilon^{mnlp}$, that is a pseudo-scalar,
on the torus.  Therefore the metric tensor, the
dilaton, the three form field strength and the
torus modulus depend only on the $x^{\alpha}$
coordinates, and they are of the above form.  
This type of dimensional reduction
picks out only the zero modes on the $T^4$ in the 
Kaluza-Klein reduction process.  Using the 
formulas given in
Appendix B for the block diagonal metric,
we get the six-dimensional action
\begin{equation}
I^{(6)} = \int d^6 x \sqrt{- g^{(6)} } (
e^{2 \psi - 2 \phi} ( R^{(6)} +
3 (D \psi )^2 - 8 D_{\alpha} \psi D^{\alpha}
\phi + 4 (D \phi )^2 ) - \frac{1}{12} H^{(6) 2} ) ,
\end{equation}
where the covariant derivative $D$ are with
respect to the six-dimensional metric.

We further compactify the six-dimensional
theory on a circle $S^1$ to obtain the
five-dimensional theory on ${\cal M}^5$.  
We again assume the maximal
symmetry on the covering space $R^1$ (one 
dimensional Euclidean Poincare symmetry) of the
$S^1$.  Thus, we assume the six-dimensional
metric of the form
\begin{equation}
g^{(6)}_{\mu \nu} = \left( \begin{array}{cc}
  g^{(5)}_{\alpha \beta}+e^{\psi_1} A_{\alpha}A_{\beta}  &   
   e^{\psi_1} A_{\alpha}   \\
    e^{\psi_1} A_{\beta}    &  e^{\psi_1}   \end{array}
                      \right)  ,
\end{equation}
where all the entries are independent of the
the position $\theta$ on the circle.  The scalar
field $\psi_1$ is the circle modulus.  
The six-dimensional three form field 
strength $H^{(6)}$ decomposes
into the five-dimensional three form field 
strength $H^{(5)}_{\alpha \beta \gamma} =  
H^{(6)}_{\alpha \beta \gamma}$
and the two form field strength
$F_{2 \alpha \beta} = H^{(6)}_{\alpha \beta \theta}$.
Using the
formulas summarized in the Appendix A, we retain
only the zero modes on the $S^1$ to obtain the 
dimensionally reduced five-dimensional action
\begin{equation}
I^{(5)} = \int d^5 x \sqrt{- g^{(5)} } (
e^{2 \psi +\psi_1 / 2  - 2 \phi} ( R^{(5)} 
+ 2D_{\alpha} \psi D^{\alpha} \psi_1
- 2D_{\alpha} \phi D^{\alpha} \psi_1
\label{5action}
\end{equation}
\[ - 8D_{\alpha} \phi D^{\alpha} \psi
+ 3 (D \psi )^2 + 4 (D \phi )^2 ) 
- \frac{1}{12} H^{\prime 2} 
- \frac{1}{4} e^{- \psi_1} 
F_{2 \alpha \beta} F_2^{\alpha \beta}
- \frac{1}{4} e^{ -2 \phi + \psi_1}
F_{\alpha \beta} F^{\alpha \beta} ) . \]
The covariant derivative $D$ and the summation 
of the indices, that run from zero to four,
are on  the five-dimensional 
manifold.  
The two form field strength $F = d A$ denotes
the Kaluza-Klein $U(1)$ gauge field that originates
from the momentum along $S^1$.
In Eq. (\ref{5action}), the three form 
$H^{\prime}$ is given by 
\[ H^{\prime} = H^{(5)} - A \wedge F_2 .\]
Via the construction so far, we obtained a five
dimensional theory on the external space-time
${\cal M}^5$ described by the action (\ref{5action}).
The compact submanifold $S^1 \times T^4$ 
of the ten dimensional space-time plays the role of the
internal space-time.  For this interpretation to be meaningful,
we should require that the circle modulus 
$\exp ( \psi_1  )$ and the four-torus modulus $\exp (\psi ) $
be bounded from the above for all points on the external manifold 
${\cal M}^5$.

The main focus of this note is to investigate the
$s$-wave sector of the five-dimensional action
Eq. (\ref{5action}).  Thus, we impose the rotational
symmetry and consider the five dimensional manifold
of the form ${\cal M}^5 = {\cal M}^2 \times S^3$.  
The compatible metric
can be written as
\begin{equation}
ds^2 = g^{(2)}_{\alpha \beta} dx^{\alpha} dx^{\beta}
+ e^{- 2 \psi_2} d \Omega^{(3)}
\end{equation}
where the two-dimensional metric 
$g^{(2)}_{\alpha \beta}$
on the ${\cal M}^2$ and the radius of the 
sphere $\exp ( - \psi_2 )$
depend only on the coordinates $x^{\alpha}$ on the 
${\cal M}^2$.  The metric $d \Omega^{(3)}$ is the metric
on the unit $S^3$ with three angle coordinates 
$\theta_1$, $\theta_2$ and $\theta_3$.  To make
the five dimensional manifold ${\cal M}^5$ non-compact,
the radius of the $S^3$ should not 
have a finite upper bound on the ${\cal M}^2$.  
Imposing the rotational symmetry 
requires that the scalar fields in the
five dimensional action (\ref{5action}) depend
only on the $x^{\alpha}$ coordinates.
In addition, it is necessary that the gauge invariant  
field strengths be the scalars on the $S^3$ or be 
proportional to the pseudo-scalar
$\epsilon_{\theta_1 \theta_2 \theta_3}$, the
volume form on the unit three-sphere.  Thus, the 
Kaluza-Klein two form field strength $F$ has 
non-vanishing components $F_{\alpha \beta}$,
and the other components that have indices 
along the $S^3$ directions vanish.  The similar
result holds for the two form field strength 
$F_{2}$.  Thus in the five-dimensional action
(\ref{5action}), $H^{ \prime} = H^{(5)}$ since 
we can always choose the vector potential 
$A$ in such a way that $A = A_{\alpha} dx^{\alpha}$.
For the three form field strength
$H^{(5)}$, the components 
$H^{(5)}_{\alpha \beta \gamma}$ 
on the ${\cal M}^2$ vanish due to the antisymmetry of
indices.  The only non-vanishing component for 
$H^{(5)}$ is $H^{(5)}_{\theta_1 \theta_2 \theta_3}
= H ^{(2)} \epsilon_{\theta_1 \theta_2 \theta_3}$,
where $H^{(2)}$ is the zero form field strength
on the ${\cal M}^2$.    The five dimensional
equations of motion for the
$H^{(5)}$ are identically satisfied by the
imposition of the rotational symmetry.  From the
point of view of the ${\cal M}^2$, the equations of
motion for the zero form field strength are vacuous.
Still, the five dimensional Bianchi identity 
$d H^{(5)} = 0$ implies two 
equations $\partial_{\alpha} H^{(2)} = 0$.

After the dimensional reduction to the ${\cal M}^2$
using the formulas from the Appendix B while recalling
that the $S^3$ has intrinsic curvature,
we find the following two dimensional action
\begin{equation}
I = \int d^2 x \sqrt{-g} (  e^{- 2 \bar{\phi} }
( R + 6 e^{ k \bar{\phi}} + (\frac{16}{3} - 2k )
( D \bar{\phi} )^2 
\label{dgra}
\end{equation} 
\[ - \frac{1}{2} (Df)^2 - \frac{1}{2} (Df_1 )^2 
- \frac{1}{2} (Df_2 )^2
- \frac{1}{2} D_{\alpha} f_1 D^{\alpha} f_2 )  \]
\[ - \frac{1}{4} e^{-\sqrt{2} f_1 - (k + 2/3) \bar{\phi} } F_1^2  
    - \frac{1}{4} e^{-\sqrt{2} f_2 - (k+ 2/3) \bar{\phi}} F_2^2
    - \frac{1}{4} e^{ \sqrt{2} (f_1 + f_2 ) - (k + 2 / 3) \bar{\phi}}
     F^2 ) . \]
The covariant derivatives are taken
with respect to
the Weyl-rescaled two dimensional metric
$g_{\alpha \beta} = \exp (  2 \psi_2 
- k \bar{\phi} ) g^{(2)}_{\alpha \beta} $.  
We introduced four scalar fields $\bar{\phi}$,
$f_1$, $f_2$ and $f$ via
\begin{equation}
-2 \bar{\phi} = -2 \phi + \frac{1}{2} \psi_1 
 -3 \psi_2  + 2 \psi  \ , \   f_1
= - \frac{\sqrt{2}}{3} \phi + \frac{\sqrt{2}}{3} \psi_1 
 + \frac{4\sqrt{2}}{3} \psi 
\end{equation}
\[ f_2  
= - \frac{\sqrt{2}}{3} \phi + \frac{\sqrt{2}}{3} \psi_1 
- \frac{2 \sqrt{2}}{3} \psi \ , \  
f = \sqrt{2} (\psi - \phi  ) ,\]
which can be inverted to give
\begin{equation}
\psi = \frac{1}{2 \sqrt{2}} (f_1 - f_2 ) \ , \
\phi = \frac{1}{2 \sqrt{2}} (f_1 - f_2 ) 
- \frac{1}{\sqrt{2}} f 
\label{sr}
\end{equation}
\[ \psi_1 = \frac{3}{2 \sqrt{2}} ( f_1 + f_2 ) 
- \frac{1}{\sqrt{2}} f  \ , \
\psi_2 = \frac{2}{3} \bar{\phi} + \frac{1}{4 \sqrt{2}}
( f_1 + f_2 ) + \frac{1}{2 \sqrt{2}} f .\]
In addition, we used an equivalent, but more
convenient description of the $H^{(2)}$ field
by taking the two-dimensional Hodge dual of 
the zero-form field strength 
$H^{(2)}$.  Namely, we take 
the dual transformation via
\begin{equation}
F_{1 \alpha \beta} = 
e^{\sqrt{2} f_1 + (k+2/3) \bar{\phi}} H^{(2)} 
\varepsilon_{\alpha \beta}  
\rightarrow
H^{(2)} = -\frac{1}{2} 
e^{-\sqrt{2} f_1 - (k+2/3) \bar{\phi}} \varepsilon^{\alpha \beta}
F_{1 \alpha \beta}
\label{2dual}
\end{equation}
to introduce an extra two-form field 
strength $F_{1 \alpha \beta}$, 
i.e., a $U(1)$ gauge field, which carries 
five-brane charges.
At the action level, the term 
\[
\int d^2 x \sqrt{-g} \frac{(-1)}{2} 
e^{\sqrt{2} f_1 + (k + 2/3) \bar{\phi}}  H^{(2) 2} 
\]
was changed to 
\begin{equation}
I = \int d^2 x \sqrt{-g} \frac{(-1)}{4}
e^{- \sqrt{2} f_1 - (k+2/3) \bar{\phi}} 
g^{(2) \alpha \beta} g^{(2)\gamma \delta} 
F_{1 \alpha \gamma} F_{1 \beta \delta} 
\end{equation}
to implement the Hodge duality.  This change
ensures that the equations of motion on the ${\cal M}^2$ 
are equivalent in both cases.  
We notice that the Bianchi identities for the
zero-form field strength become the equations
of motion in the dual picture.  Just as the 
equations of motion for the zero-form field 
strength are vacuous, the Bianchi identities
for the two-form field strength in two dimensions
are vacuous.  

 We observe that the action Eq. (\ref{dgra}) is a
two dimensional dilaton gravity with a particular set of
field contents.  The circle modulus, the four-torus modulus,
the $S^3$ radius and the ten-dimensional dilaton field
combine to give four scalar fields on ${\cal M}^2$.     
There are three $U(1)$ field strengths, $F_1$, $F_2$
and $F$, that are produced by the five-brane
charges, one-brane charges and the Kaluza-Klein
charges, respectively.   On ${\cal M}^5$, to get the
charge of the $H^{(5)}_{\theta_1 \theta_2 \theta_3}$
that gives $F_1$ via Eq. (\ref{2dual}), we can directly integrate 
the field strength on the $S^3$.  Thus, its charges
are magnetic-like and its ten-dimensional dual field
strength has the dual gauge field 
$\tilde{A}_{\alpha \theta x^6 x^7 x^8 x^9}$.  
This naturally couples to the five D-branes wrapped 
on the $S^1 \times T^4$.  To obtain the charge of the
$F_{2 \alpha \beta}$, we have to integrate the 
five dimensional dual of $F_2$ on the $S^3$.  This
means that its charges are electric-like. 
The ten-dimensional gauge field for the $F_2$ field
then satisfies  
\[ F_{2 \alpha \beta} = 
H^{(10)}_{\alpha \beta \theta} = \partial_{\alpha}
A_{\beta \theta} - \partial_{\beta} A_{\alpha \theta}  . \]
We see that $A_{\alpha \theta}$ naturally couples to
the one D-branes wrapped on the circle $S^1$.
 
The ten dimensional symmetries
are still present as internal symmetries of the action
Eq.(\ref{dgra}).  For example, the ten dimensional Hodge
duality that exchanges one-branes and five-branes
corresponds to the symmetry of Eq. (\ref{dgra}) under
the transformation
\[ F_1 \rightarrow F_2 \ , \ F_2 \rightarrow F_1 
   \ , \ f_1 \rightarrow f_2 \ , \ f_2 \rightarrow f_1 . \] 
When $f = 0$, which  
holds for solutions without naked 
singularities that  we will explain in the later section,
 Eq. (\ref{sr}) shows
that the above transformation inverts the $T^4$
modulus and the sign of the ten-dimensional 
dilaton field, a characteristic of the $U$-duality.

\section{General static solutions}

To get the general static solutions of the bosonic
$s$-wave sector of the type IIB supergravity
compactified on $S^1 \times T^4$, we need to only
consider the two dimensional dilaton gravity described
by Eq. (\ref{dgra}).  The general static solutions of
the similar dilaton gravity theories were obtained in
Ref. \cite{kp}.  By using a variant of the arguments
given in that paper, we will first obtain the general
static solutions on a local coordinate patch.  In 
particular, we make no assumptions about
the global and asymptotic structure of the space-time.
The discussions on the properties of the solutions
will follow.
 
\subsection{Static equations of motion}

For the description of the space-time geometry 
on a local coordinate patch in ${\cal M}^2$, we choose 
to use a conformal gauge for the two dimensional
metric 
\begin{equation}
ds^2 = g_{\alpha \beta} dx^{\alpha} dx^{\beta} 
= - \exp ( 2 \rho ) dx^+ dx^-  .
\end{equation}
We also choose the number $k= 8/3$ in the action 
Eq. (\ref{dgra}) to cancel the kinetic term for the
two dimensional dilaton field $\bar{\phi}$.
A different choice of the value of $k$ is 
related to this choice by a Weyl rescaling of the
two dimensional metric.

Under this gauge choice, the equations of motion
from Eq. (\ref{dgra}) are given as follows.
By varying the action with respect to the 
two-dimensional dilaton field $\bar{\phi}$,
we have 
\begin{equation}
4 \partial_+ \partial_- \rho 
- e^{2 \rho} \Omega^{- 4/3} 
+  \partial_+ f \partial_- f
+  \partial_+ f_1 \partial_- f_1
+  \partial_+ f_2 \partial_- f_2
+ \frac{1}{2} \partial_+ f_1 \partial_- f_2
\label{eom1}
\end{equation}
\[ + \frac{1}{2} \partial_+ f_2 \partial_- f_1
 + \frac{5}{3} e^{-2 \rho} \Omega^{2/3} (
e^{-\sqrt{2} f_1} F_{1 +-}^2 
+ e^{-\sqrt{2} f_1} F_{2 +-}^2
+ e^{\sqrt{2} ( f_1 + f_2 )} F_{+-}^2 ) = 0 \]
where we introduce $\Omega = \exp ( -2 \bar{\phi} )$.
Similarly for the conformal factor $\rho$ of 
the metric, we get
\begin{equation}
2 \partial_- \partial_+ \Omega + 3e^{2 \rho}
\Omega^{-1 /3} 
-  e^{-2 \rho} \Omega^{5/3} (
e^{-\sqrt{2} f_1} F_{1 +-}^2 
+ e^{-\sqrt{2} f_1} F_{2 +-}^2
+ e^{\sqrt{2} ( f_1 + f_2 )} F_{+-}^2 ) = 0 .
\end{equation}
The equations of motion for the scalar fields become
\begin{equation}
\partial_+ ( \Omega \partial_- f )
+ \partial_- ( \Omega \partial_+ f ) = 0 ,
\end{equation}
\begin{equation}
2 \partial_+ ( \Omega \partial_- f_1 )
+ 2 \partial_- ( \Omega \partial_+ f_1 )
+ \partial_+ ( \Omega \partial_- f_2 )
+ \partial_- ( \Omega \partial_+ f_2 )
\end{equation}
\[ + 2\sqrt{2} e^{-2 \rho} \Omega^{5/3}
( e^{- \sqrt{2} f_1} F_{1 +-}^2 
 - e^{\sqrt{2} (f_1 + f_2 ) }F_{+-}^2 ) = 0,
\]
and
\begin{equation}
2 \partial_+ ( \Omega \partial_- f_2 )
+ 2 \partial_- ( \Omega \partial_+ f_2 )
+ \partial_+ ( \Omega \partial_- f_1 )
+ \partial_- ( \Omega \partial_+ f_1 ) 
\end{equation}
\[ + 2\sqrt{2} e^{-2 \rho} \Omega^{5/3}
( e^{- \sqrt{2} f_2} F_{1 +-}^2 
 - e^{\sqrt{2} (f_1 + f_2 ) }F_{+-}^2  ) = 0 .
\]
We also have equations for the three $U(1)$
gauge fields
\begin{equation}
\partial_+ ( e^{-\sqrt{2} f_1 -2 \rho} 
\Omega^{5/3} F_{1 +-} )
= \partial_- ( e^{-\sqrt{2} f_1 -2 \rho} 
\Omega^{5/3} F_{1 +-} ) = 0 ,
\end{equation}
\begin{equation}
\partial_+ ( e^{-\sqrt{2} f_2 -2 \rho} 
\Omega^{5/3} F_{2 +-} )
= \partial_- ( e^{-\sqrt{2} f_2 -2 \rho} 
\Omega^{5/3} F_{2 +-} ) = 0 ,
\end{equation}
and
\begin{equation}
\partial_+ ( e^{\sqrt{2} (f_1 + f_2 ) -2 \rho} 
\Omega^{5/3} F_{+-} )
= \partial_- ( e^{\sqrt{2} (f_1 + f_2 ) -2 \rho} 
\Omega^{5/3} F_{+-} ) = 0 .
\label{eom2}
\end{equation}
These equations of motion, Eqs. (\ref{eom1}) -
(\ref{eom2}), should be supplemented with the
gauge constraints produced by the choice of a
conformal gauge
\begin{equation}
T_{++} = T_{--} = 0 ,
\label{gcon}
\end{equation}
where $T_{\pm \pm}$ denotes the $\pm \pm$
components of the stress-energy tensor.

By requiring that all the fields depend only
on a space-like coordinate $x \equiv x^+ - x^-$,
we restrict our attention only to the static solutions.
We also make a convenient gauge choice for the
$U(1)$ gauge fields by introducing three functions
(in fact, three static potentials) 
$A(x)$, $A_1 (x)$, and $A_2 (x)$ by
\begin{equation}
A_{\pm} = \frac{1}{2} A (x)  \ , \ 
A_{1 \pm} = \frac{1}{2} A_1 (x) \ , \
A_{2 \pm} = \frac{1}{2} A_2 (x) ,
\end{equation}
where the gauge fields give the field strengths via 
$F_{2 +-} = \partial_+ A_{2 -} - \partial_- A_{2 +}$,
$F_{1 +-} = \partial_+ A_{1 -} - \partial_- A_{1 +}$,
and
$F_{ +-} = \partial_+ A_{-} - \partial_- A_{ +}$.
The static equations of motion are then 
summarized by a one-dimensional action
\begin{equation}
I = \int dx ( \Omega^{\prime} \rho^{\prime}
+ \frac{3}{4} e^{2 \rho} \Omega^{-1 / 3}
- \frac{1}{4} \Omega f^{\prime 2} 
-\frac{1}{4}  \Omega f_1^{\prime 2} 
- \frac{1}{4} \Omega f_2^{\prime 2}
-\frac{1}{4} \Omega f_1^{\prime} f_2^{\prime}
\label{eom3}
\end{equation}
\[  + \frac{1}{4} e^{- \sqrt{2} f_1 - 2 \rho} 
   \Omega^{5/3} A_1^{\prime 2}
     + \frac{1}{4} e^{ - \sqrt{2} f_2 - 2 \rho}
   \Omega^{5/3} A_2^{\prime 2}
      + \frac{1}{4} e^{\sqrt{2} ( f_1 + f_2 ) - 2 \rho}
    \Omega^{5/3} A^{\prime 2}   )  , \]
while the gauge constraints Eq.(\ref{gcon}) become
a single condition
\begin{equation}
\Omega^{\prime \prime} - 2 \rho^{\prime} 
\Omega^{\prime} + \frac{1}{2} \Omega f^{\prime 2}
+ \frac{1}{2} \Omega f_1^{\prime 2}
+ \frac{1}{2} \Omega f_2^{\prime 2}
+ \frac{1}{2} \Omega f_1^{\prime} f_2^{\prime} = 0 . 
\label{eom4}
\end{equation} 
The prime denotes the differentiation with respect
to $x = x^+ - x^-$.
Our task from now on is to solve the equations
of motion from the action Eq.(\ref{eom3}) under the
gauge constraint Eq.(\ref{eom4}).

\subsection{Symmetries and general static solutions}

The main virtue of writing down the one dimensional
action Eq. (\ref{eom3}) is that it enables us to clearly
identify the symmetries of our problem.  We find
that there are eight symmetries
\begin{eqnarray*}
(a) \ \ &&f \rightarrow f + \alpha, \\
(b) \ \ &&f_1 \rightarrow f_1 + \alpha, \ A_1 \rightarrow 
  A_1 e^{\alpha / \sqrt{2} }, \ A \rightarrow A 
  e^{- \alpha / \sqrt{2} }, \\
(c) \ \ &&f_2 \rightarrow f_2 + \alpha, \ A_2 \rightarrow 
  A_2 e^{\alpha / \sqrt{2} }, \ A \rightarrow A 
  e^{- \alpha / \sqrt{2} }, \\
(d) \ \ &&A_1 \rightarrow A_1 + \alpha,\\
(e) \ \ &&A_2 \rightarrow A_2 + \alpha,\\
(f) \ \ &&A \rightarrow A + \alpha,\\
(g) \ \ &&x \rightarrow x + \alpha,\\
(h) \ \ &&x \rightarrow e^{\alpha}  x, \ 
     \Omega \rightarrow e^{\alpha} \Omega, \
     e^{2 \rho} \rightarrow e^{ -2 \alpha / 3} 
     e^{2 \rho}, \
     A \rightarrow e^{ - 2 \alpha /3 } A, \
     A_1 \rightarrow e^{-2 \alpha /3 } A_1, \
     A_2 \rightarrow e^{-2 \alpha /3 } A_2,
\end{eqnarray*}
where $\alpha$ is an arbitrary real parameter
of each continuous transformation.  Due to these
eight symmetries for the eight fields in our problem, 
it is possible to integrate the coupled second order 
differential equations once to get coupled first
order differential equations.  They
can be recast in the form of the Noether
charge expressions
\begin{eqnarray}
f_0 & = &\Omega f^{\prime}  ,
\label{f0} \\
f_{10} & = & \Omega f_1^{\prime} 
+ \frac{1}{2} \Omega f_2^{\prime} 
- \frac{1}{\sqrt{2}} e^{- \sqrt{2} f_1 -2 \rho }
\Omega^{5/3} A_1 A_1^{\prime} 
+ \frac{1}{\sqrt{2}} 
e^{ \sqrt{2} (f_1 + f_2 ) -2 \rho} \Omega^{5/3}
A A^{\prime} , 
\label{f10} \\
f_{20} & = & \Omega f_2^{\prime} 
+ \frac{1}{2} \Omega f_1^{\prime} 
- \frac{1}{\sqrt{2}} e^{- \sqrt{2} f_2 -2 \rho }
\Omega^{5/3} A_2 A_2^{\prime} 
+ \frac{1}{\sqrt{2}} 
e^{ \sqrt{2} (f_1 + f_2 ) -2 \rho} \Omega^{5/3}
A A^{\prime} , 
\label{f20} \\
Q_1  & =&  e^{- \sqrt{2} f_1 -2 \rho} 
\Omega^{5/3}  A_1^{\prime}, 
\label{q1} \\
Q_2  & =&  e^{- \sqrt{2} f_2 -2 \rho} 
\Omega^{5/3}  A_2^{\prime}, 
\label{q2} \\
Q  & =&  e^{\sqrt{2} (f_1 + f_2 ) -2 \rho} 
\Omega^{5/3}  A^{\prime}, 
\label{q} \\
c_0 & = &\Omega^{\prime} \rho^{\prime} 
- \frac{1}{4}\Omega  f^{\prime 2} 
- \frac{1}{4}\Omega  f_1^{\prime 2} 
- \frac{1}{4}\Omega  f_2^{\prime 2} 
- \frac{1}{4}\Omega  f_1^{\prime} f_2^{\prime} 
+ \frac{1}{4}e^{- \sqrt{2} f_1 -2\rho} \Omega^{5/3} A_1^{\prime 2}
\nonumber \\
& & + \frac{1}{4}e^{- \sqrt{2} f_2 -2\rho} \Omega^{5/3} A_2^{\prime 2}
+ \frac{1}{4}e^{\sqrt{2} (f_1 +f_2 )-2\rho} \Omega^{5/3} A^{\prime 2}
 - \frac{3}{4} e^{2 \rho} \Omega^{-1/3}, 
\label{c0} \\
s + c_0  x & = &-\frac{1}{3} \Omega^{\prime}
+ \rho^{\prime} \Omega 
- \frac{1}{3} e^{-\sqrt{2} f_1 -2 \rho} \Omega^{5/3}
  A_1 A_1^{\prime}
\nonumber \\
& & - \frac{1}{3} e^{-\sqrt{2} f_2 -2 \rho} \Omega^{5/3}
  A_2 A_2^{\prime}
- \frac{1}{3} e^{\sqrt{2} (f_1 +f_2 )-2 \rho} \Omega^{5/3}
  A A^{\prime} ,
\label{s}
\end{eqnarray}
corresponding to each symmetry.  The gauge constraint 
Eq. (\ref{eom4}) implies $c_0 = 0$, thereby reducing the
number of constants of motion from eight to seven.

In the Appendix C, we exactly solve the above 
differential equations.  The results are as follows.
\begin{eqnarray}
f(A)          &=& f_0I(A)+f_1   \\
Q_1^2 e^{2 f_a (A)}  &=& \frac{Q}{P(A)}
    \frac{c_1}{\sinh^2 \left[\sqrt{c_1}( I(A) + \tilde{c}_1 ) \right]}  \\
Q_2^2 e^{2 f_b (A)}  &=& \frac{Q}{P(A)}
    \frac{c_2}{\sinh^2 \left[\sqrt{c_2}( I(A) + \tilde{c}_2 ) \right]}  \\
\bar{\Omega}^{4/3}(A) &=& \frac{Q}{P(A)}
    \frac{D_2}{\sinh^2 \left[\sqrt{D_2}( I(A) + \tilde{c} ) \right]}  \\
Q_1 A_1 (A)   &=& -\sqrt{2}f_{10}-\bar{s} - \sqrt{c_1 }\coth
    \left[\sqrt{c_1} (I(A) + \tilde{c}_1 ) \right]  \\
Q_2 A_2 (A)   &=& -\sqrt{2}f_{20}-\bar{s} - \sqrt{c_2 }\coth
    \left[\sqrt{c_2} (I(A) + \tilde{c}_2 ) \right] \\
e^{2\bar{\rho} (A) }  &=&\frac{P(A)}{Q} \bar{\Omega}^{2/3}(A) \\
x-x_0         &=& \int \frac{\Omega(A)}{P(A)}dA
\end{eqnarray}
where
\[ f_a =(2f_1+f_2)/\sqrt{2},~~~ f_b =(f_1+2f_2)/\sqrt{2},\]
\[ e^{2\bar{\rho}}=e^{-(f_a + f_b )/3}e^{2\rho},
 \ \bar{\Omega}=e^{(f_a +f_b )/2}\Omega,
\]
\[P(A)=QA^2+2\bar{s}A+c,\]
\[\bar{s}=s-\frac{\sqrt{2}}{3}(f_{10}+f_{20}),
 \ D_2=\frac{1}{3}\left(g_0^2+\bar{s}^2-Qc+c_1 + c_2 \right),\]
and we introduce a function
\[ I(A) = \int \frac{dA}{P(A)}
= \frac{1}{2 \sqrt{\bar{s}^2 - Qc}} \ln 
  \left| \frac{A-A_+}{A-A_-} \right| .\]
Here $A_{\pm}$ denotes the two solutions of the
quadratic equation $P(A) = 0$ given by
\[ A_{\pm} = \frac{ - \bar{s} \pm \sqrt{\bar{s}^2 - Qc}}
    {Q}  .\]
The parameters $f_1$, $c_1$, $c_2$, $\tilde{c}_1$, 
$\tilde{c}_2$, $\tilde{c}$, $c$ and $x_0$ are 
eight constants of integration that are produced when
we solve the coupled first order differential equations
involving eight fields.  Combined with seven non-vanishing
Noether charges, we find that the general static solutions
are labeled by fifteen parameters.

\subsection{Properties of general static solutions}

The general static solutions we obtained in the 
previous subsection
contain fifteen parameters.   Generically,
the local structure of the solution space can be 
understood as follows.  We have three gauge fields
that are produced by the one-brane charge 
$Q_{1-brane}$, the 
five-brane charge
$Q_{5-brane}$ and the Kaluza-Klein charge $n$.  The asymptotic
values of the gauge potentials provide additional three
parameter $A_{\infty}$, $A_{1 \infty}$, and $A_{2 \infty}$,
the local variation of which near a generic point in
the solution space can not be physically observed, since 
they are not gauge invariant.  
In addition, we have three scalar fields 
that correspond to the circle modulus
$\exp (\psi_1 )$, the four-torus modulus $\exp (\psi )$,
and the dilaton field $\phi$.  The asymptotic values of these
three scalar fields at the spatial infinity (when they exist)
specify the asymptotic size of the circle 
$\exp (\psi_{1 \infty} /2 )$, the four-torus volume 
$\exp (\psi_{ \infty} /2 )$,
and the the asymptotic dilaton coupling constant
$\exp ( \phi_{\infty} )$.  The three scalar fields
also carry charges $Q_{\psi_1}$, $Q_{\psi}$ and
$Q_{\phi}$.  The gravity sector provides
additional three parameters; the black hole mass $M$
(when it exists, or the mass parameter when it 
does not exist), the length-scale choice, and the
reference time choice.  The length-scale choice is
fixed when we require that 
the five dimensional metric
become the standard flat Minkowskian when all the 
other charges are set to zero.  The reference time
choice is an isometry and it is an arbitrary
number, as long as we consider the static solutions.

However, the global structure of the solution space
exhibits much richer features.  To see this clearly,
we write down the expressions for the circle modulus, 
the four-torus modulus, the radius of the $S^3$, and 
the ten dimensional
dilaton field.  For the circle modulus, we have
\begin{equation}
e^{{\psi}_1}=\left|\frac{Q \sqrt{c_1 c_2}}
{ Q_1 Q_2} \right|^{1/2} 
\frac{ e^{-(f_0 I(A)+f_1)/\sqrt{2}}}
{\left| P(A) \sinh \left[ \sqrt{c_2} (I(A) +
\tilde{c}_2 ) \right] \sinh \left [
\sqrt{c_1} (I(A) + \tilde{c}_1 ) \right ]
  \right|^{1/2}}  ,
\label{sm} 
\end{equation}
the four-torus modulus is given by
\begin{equation}
e^{\psi} = 
\left| \frac{Q_2 \sqrt{c_1}}{Q_1 \sqrt{c_2}}
\frac{\sinh \left[ \sqrt{c_2} (I(A) +
\tilde{c}_2 ) \right]}{\sinh \left [
\sqrt{c_1} (I(A) + \tilde{c}_1 ) \right ]
  } \right|^{1/2} , 
\label{tm}
\end{equation}
the square of the $S^3$ radius becomes
\begin{equation}
e^{-2\psi_2} = \left| \frac{D_2 Q_1 Q_2}{\sqrt{c_1 c_2}}
\frac{\sinh \left[ \sqrt{c_2} (I(A) +
\tilde{c}_2 ) \right] \sinh \left [
\sqrt{c_1} (I(A) + \tilde{c}_1 ) \right ] } 
{ \sinh^2 \left[\sqrt{D_2}(I(A)+\tilde{c}) \right]
   } \right|^{1/2} 
e^{-(f_0 I(A)+f_1)/\sqrt{2}}  ,
\label{sphr}
\end{equation}
and we have the ten-dimensional dilaton expression
\begin{equation}
e^{-2\phi} = \left| \frac{ Q_1 \sqrt{c_2}}
        {Q_2 \sqrt{c_1}}
        \frac{\sinh \left[ \sqrt{c_1} 
        (I(A) + \tilde{c}_1 )
        \right]}{\sinh \left[ \sqrt{c_2} 
        (I(A) + \tilde{c}_2 )
        \right]}  \right|
e^{\sqrt{2}(f_0 I(A)+f_1)} .
\label{td}
\end{equation}
The two dimensional metric becomes
\begin{equation}
ds^2 = g^{(2)}_{\alpha \beta} dx^{\alpha} dx^{\beta}
= - e^{2 \rho- 2 \psi_2 + 8 \bar{\phi} / 3} dx^+ dx^-
= - \frac{P(A)}{Q} e^{\psi_1} dx^+ dx^-.
\label{2dm} 
\end{equation}
We immediately see that the
qualitative behaviors of the solutions change when we change
any sign(s) of $D_2$, $c_1$ or $c_2$, or when we make any
of them become zero.  Keeping this fact in mind,
we restrict out attention only to the case when
$D_2 > 0$, $c_1 > 0$ and $c_2 > 0$, which we will see
shortly corresponds to the cases considered in     
Ref. \cite{hms}.  We furthermore assume that the 
quadratic equation $P(A) = 0$ for $A$ has two real 
solutions $A_{\pm}$, i.e., $\bar{s}^2 - Qc \ge 0$.  This 
condition implies that $3 D_2 - c_1 - c_2 \ge 0$.
Under these conditions, we look for the five dimensional
solutions.  The five dimensional asymptotic spatial infinities
are the points on ${\cal M}^2$, near of which the $S^3$ radius
Eq. (\ref{sphr}) tends to diverge.  At the same time,
the circle modulus Eq. (\ref{sm}) and the four-torus
modulus should remain bounded from the above.  
Since 
\[ I(A) =   
\frac{1}{2 \sqrt{\bar{s}^2 - Qc}} \ln \left| 
\frac{A - A_+}{A-A_-}\right| , \]
the potentially divergent points for the
$S^3$ radius are $A = A_{\pm}$ and the two values
$A = A^{\pm}_{\infty}$ satisfying
\begin{equation}
 I(A^{\pm}_{\infty} ) + \tilde{c} = 0 .
\label{infi}
\end{equation}
Assuming that the circle modulus are bounded from the above
near the asymptotic spatial infinities and, further
assuming that they form a time-like curve, Eq. (\ref{2dm})
shows that $A = A_{\pm}$ can not be the asymptotic
spatial infinities.  

     Depending on the sign of 
$\tilde{c}$, we have to consider the three cases.
If $\tilde{c} > 0$, we find that $A_{\pm} < A^+_{\infty}$
and $min (A_- , A_+ ) < A^-_{\infty} < max (A_- , A_+ )$.
If $\tilde{c} = 0$, we find that $A^{\pm}_{\infty} = \pm
\infty$.  If $\tilde{c} < 0$, we have  $A^-_{\infty} < A_{\pm}$
and $min (A_- , A_+ ) < A^+_{\infty} < max (A_- , A_+ )$.
One can verify that the first and the third case describes
the identical situation.  Thus, we concentrate on the
first case and let the value of $A$ decrease starting from
$A = A^+_{\infty}$, i.e., the asymptotic spatial 
infinities.  Since the circle modulus and the four-torus
modulus should remain finite until we hit the null curve
(outer event horizon), $A= max(A_- , A_+ )$ as can be
seen from Eq. (\ref{2dm}) , we have to impose $\tilde{c}
> \tilde{c}_1$ and $\tilde{c} > \tilde{c}_2$.  The condition 
that the moduli fields do
not blow up near the potential outer horizon
gives us the three conditions
\begin{equation}
f_0 = 0 \ , \ D_2 = c_1 \ , \  D_2 = c_2 
\ \rightarrow \ D_2 = c_1 = c_2 =  \bar{s}^2 - Qc .
\label{bhc}
\end{equation} 
If these conditions are not met, we have naked singularities
at the would-be outer horizon.  Thus, if the charges
of the scalar moduli fields are not set to a certain value,
as is familiar from the four dimensional stringy black holes
\cite{horowitz}, we have naked singularities.  This behavior
is consistent with the no-hair property and the familiar
five dimensional black hole 
solutions \cite{cvetic2} \cite{hms} share this property.

To better understand our solutions, we recast them
in a radial gauge.  We focus on the case when
$A_- \le A_+ < 0 $ and, furthermore, to get the description
of the space-time outside the outer event horizon, we
consider the range of $A > A_+$.  Under these 
conditions, we note that $\sqrt{D_2} (I(A) + \tilde{c})
\le 0$. A convenient radial coordinate $r$ choice is
\begin{equation}
r^2 = - \sqrt{\frac{D_2 Q}{c}} 
\left[ \coth ( \sqrt{D_2} (I(A) + \tilde{c}) ) 
-1 \right] ,
\end{equation}
which becomes 
\[
( 1- \frac{ r_0^2}{r^2} )^{\sqrt{\frac{\bar{s}^2 -Qc}
{D_2}}} = e^{2 \tilde{c} \sqrt{\bar{s}^2 - Qc}}
\left( \frac{A- A_+}{A- A_-} \right) ,
\]
where we define
\[ r_0^2 = 2 \sqrt{\frac{D_2 Q}{c}} . \]
As can be seen from the above expression, the range
of $\tilde{c}$ is restricted to be $\tilde{c} \ge 0$
if we require that the range $A_+ < A <
A^+_{\infty}$ maps into $r_0 < r < \infty$.  The natural
time coordinate $t$ in our context is
$t = (x^+ + x^- )$, for our fields depend
only on a space-like coordinate $x = x^+ - x^-$.   

In terms of these $(t,r)$ coordinates, the 
ten dimensional metric
in the radial gauge and the dilaton field 
can be written as follows.
\begin{eqnarray}
ds^2 &=&
Z_{\phi} Z_1^{-1/2}Z_5^{-1/2} f_1 (r) \left[
- \beta^2 dt^2 + d\theta^2 + f_2(r) 
\left(\beta \cosh \sigma dt + \sinh \sigma d\theta
\right)^2 \right] \nonumber \\
&& +Z_1^{1/2}Z_5^{-1/2}dy^m dy^m 
+ Z_{\phi} Z_1^{1/2}Z_5^{1/2}\left(Z_0^{-1} dr^2
+ r^2 d\Omega^{(3)} \right),
\label{wow}
\end{eqnarray}
\begin{equation}
e^{-2 \phi} = Z_{\phi}^{-2}
 \frac{Z_5}{Z_1} ,  
\end{equation} 
where we introduce six functions
\[ f_1 (r) = \frac{\sqrt{Qc}}{\bar{s} +
   \sqrt{\bar{s}^2 - Qc}} e^{\tilde{c} 
   \sqrt{\bar{s}^2 - Qc}}
   \left( 1 - \frac{r_0^2}{r^2}
   \right)^{(1- \sqrt{(\bar{s}^2 - Qc) / D_2 } ) /2 } ,\]
\[ f_2 (r) = 1 -
   \frac{\bar{s} + \sqrt{\bar{s}^2  - Qc}}
     {\bar{s} - \sqrt{\bar{s}^2  - Qc}}
   e^{ -2 \tilde{c} \sqrt{\bar{s}^2 - Qc}}
   \left( 1- \frac{r_0^2 }{r^2} 
   \right)^{\sqrt{(\bar{s}^2 - Qc) / D_2 } } , \]
\[ Z_0 = 1 - \frac{r_0^2}{r^2} , \]
\[ Z_1 = \sqrt{\frac{c Q_2^2 }{4 c_2 Q}}
   \left[ e^{\sqrt{c_2} ( \tilde{c} - \tilde{c}_2 )}
   \left( 1 - \frac{r_0^2}{r^2} \right)^{(1 - 
   \sqrt{c_2 / D_2 } ) /2 }
   -e^{- \sqrt{c_2} ( \tilde{c} - \tilde{c}_2 )}
   \left( 1 - \frac{r_0^2}{r^2} \right)^{(1 + 
   \sqrt{c_2 / D_2 } ) /2 } \right] , \]
\[ Z_5 = \sqrt{\frac{c Q_1^2 }{4 c_1 Q}}
   \left[ e^{\sqrt{c_1} ( \tilde{c} - \tilde{c}_1 )}
   \left( 1 - \frac{r_0^2}{r^2} \right)^{(1 - 
   \sqrt{c_1 / D_2 } ) /2 }
   -e^{- \sqrt{c_1} ( \tilde{c} - \tilde{c}_1 )}
   \left( 1 - \frac{r_0^2}{r^2} \right)^{(1 + 
   \sqrt{c_1 / D_2 } ) /2 } \right] , \]
\[ Z_{\phi} = e^{( f_0 \tilde{c} - f_1 ) /\sqrt{2}}
   \left( 1 - \frac{r_0^2}{r^2} \right)^{
   - f_0 / (2 \sqrt{2 D_2 } ) } , \]
and two parameters
\[ \beta = \sqrt{\frac{c}{4Q}} , \]
\[ \tanh \sigma = - \left(
   \frac{\bar{s} - \sqrt{\bar{s}^2 - Qc}}
     {\bar{s} + \sqrt{\bar{s}^2 - Qc}}
   \right)^{1/2}  . \]
where the parameters $\beta$ and $\sigma$ represent
the time scale and the boost parameter along the
circle $S^1$, respectively.  Generically, the space-time
described by the metric Eq.(\ref{wow}) has naked
singularities at the would-be outer horizon at $r = r_0$.
To get the black hole solutions, we have to impose
the three conditions in Eq. (\ref{bhc}).  
Under these conditions, the functions introduced
above get much simpler.  For example, the functions
$Z_1$ becomes
\begin{equation}
 Z_1 = \frac{ |Q_2|}{r_0^2}
   \left( 2 \sinh ( \sqrt{c_2} 
    ( \tilde{c} - \tilde{c}_2 )  )
   + e^{- \sqrt{c_2} ( \tilde{c} - \tilde{c}_2 )} 
    \frac{r_0^2}{r^2}  \right) , 
\label{well}
\end{equation}
which is a familiar harmonic function.

To see that our black hole solutions
indeed represent the solutions of Ref. \cite{hms},
we make further restrictions on parameters (which correspond
to the choice of the asymptotic value of the three gauge
potentials) such as
\begin{equation}
 \frac{\bar{s} + \sqrt{\bar{s}^2  - Qc}}
     {\bar{s} - \sqrt{\bar{s}^2  - Qc}} =
   e^{ 2 \tilde{c} \sqrt{\bar{s}^2 - Qc}}
\label{acon}
\end{equation}
\[  2  \frac{ |Q_2|}{r_0^2} \sinh ( \sqrt{c_2} 
    ( \tilde{c} - \tilde{c}_2 )  ) = 1 \ , \ 
2  \frac{ |Q_1|}{r_0^2} \sinh ( \sqrt{c_1} 
    ( \tilde{c} - \tilde{c}_1 )  ) = 1 . \]
In addition, we set the asymptotic value of the
ten dimensional dilaton as one by setting $f_1 = 0$,
which gives $Z_{\phi} = 1$ under the black hole 
conditions Eq. (\ref{bhc}).  In fact, under
the conditions Eqs. (\ref{bhc}) and (\ref{acon}),
we can also straightforwardly verify that 
$f_1 (r) = 1$, $f_2 (r) = r_0^2 / r^2 $ and 
\[  Z_1=1+\frac{r_1^2}{r^2} \ , \ 
  Z_5=1+\frac{r_5^2}{r^2} , \]
where
\[ r_1^2=\sqrt{Q_2^2+\frac{r_0^4}{4}}-\frac{r_0^2}{2}
 \ , \ r_5^2=\sqrt{Q_1^2+\frac{r_0^4}{4}}-\frac{r_0^2}{2}.\] 
We now explicitly see that 
Eq. (\ref{wow}) represents the six parameter solutions of
Ref. \cite{hms} in the notation of \cite{dps}.

The solution with $\tilde{c} = 0$ is our next concern. 
The similar situation in the context of Ref. \cite{kp}
results non-asymptotically flat solutions of the
type first considered in Ref. \cite{chm}.
A clear way of demonstrating this is to write down the
five dimensional metric expression.  In ten (or six)
dimensional metric expression, the static gauge 
transformation for the gauge potential $A$, i.e.,
an addition of an arbitrary number to the potential $A$, 
is part of the diffeomorphism of the ten dimensional
space-time.  This diffeomorphism becomes the internal
gauge transformation for the Kaluza-Klein gauge field
$A$ and, thus, decouples from the diffeomorphism
for the five dimensional space-time.  An explicit
calculation leads to the following five dimensional
metric expression
\begin{equation}
ds^2 = - Z_{\phi} Z_1^{-1/2}Z_5^{-1/2} f_3 (r) 
 \beta^2 dt^2
 + Z_{\phi} Z_1^{1/2}Z_5^{1/2}\left(Z_0^{-1} dr^2
+ r^2 d\Omega^{(3)} \right), 
\label{wow1}
\end{equation}
where the function $f_3 (r)$ is given by
\begin{equation}
 f_3 (r) = 2 \sqrt{\frac{\bar{s}^2 - Qc}{Qc} }
      e^{- \tilde{c} \sqrt{\bar{s}^2 - Qc}}
\frac{ \left( 1 - \frac{r_0^2}{r^2} 
      \right)^{(1 + \sqrt{(\bar{s}^2 - Qc)/D_2} \ ) /2} }
     {1- e^{-2 \tilde{c} \sqrt{\bar{s}^2 - Qc}}
     \left( 1 - \frac{r_0^2}{r^2} 
      \right)^{\sqrt{(\bar{s}^2 - Qc)/D_2}} } . 
\label{f3}
\end{equation}
We can see from Eq. (\ref{f3}) that as long as
$\tilde{c} > 0$, the function $f_3 (r)$ goes to
a finite value as $r \rightarrow \infty$.  On
the other hand, if $\tilde{c} = 0$, we find that
$f_3 (r) \rightarrow r^2 $ as
$r \rightarrow \infty$ due to the contribution
from the denominator of Eq. (\ref{f3}).
This asymptotic behavior at the spatial infinity
shows that the solutions with $\tilde{c} = 0$
are the analogues of the Chan-Horne-Mann
solutions \cite{chm} as explained in \cite{kp}.
If we indeed impose the conditions Eq. (\ref{bhc})
to consider the black-hole-like space-time structure,
we get the five dimensional metric expression
\begin{equation}
ds^2 = - U dt_1^2 + U^{-1} d \bar{r}^2 
+ \bar{r} Z_1^{1/2} Z_5^{1/2} d\Omega^{(3)} 
\end{equation}
for the $\tilde{c} = 0$ case, after the coordinate
change $\bar{r} = r^2$ and the time rescaling.
Here, the function $U$ is given by
\[ U(\bar{r} ) = 4 \bar{r}
  \left(1 - \frac{r_0^2}{\bar{r}} \right)
\frac{1}{\sqrt{(1 + r_1^2 / \bar{r})
            (1+ r_5^2 / \bar{r} ) }} .\]
Near the region $\bar{r} = r_0^2$, the space-time
looks like a black hole.  However, the asymptotic
behavior near $\bar{r} = \infty$ is quite different
from being asymptotically flat.

\section{Discussions}

Our theme in this paper is to apply the 
techniques from the lower dimensional gravity
theories to higher dimensional supergravities
by restricting our attention to the
$s$-wave sector.  In particular, we obtained the
general static solutions of the $s$-wave sector
of the type IIB supergravity on $S^1 \times T^4$.
In a sector of our solutions that reproduces the
five dimensional black hole results, we get
naked singularities and non-asymptotically flat
solutions in addition.  Still, the general 
static solutions have much richer structure;
for example, in the cases when one or more of
$D_2$, $c_1$ and $c_2$ become non-positive,
the qualitative behaviors of the solutions
change.  Even when they are all positive,
if the inequalities $\tilde{c} > \tilde{c}_1$
or $\tilde{c} > \tilde{c}_2$ get violated, 
the circle and/or the four-torus moduli blow
up (or become zero, thereby
becoming strong coupling) before we reach the 
spatial infinities.  Thus, the external 
(non-compact) dimension of the space-time depends
on the value of the constants of the integration
\footnote{We note that essentially the same 
observation was also made in \cite{hyun}, which
we received while this paper was written up.
In that reference, the 
three dimensional BTZ solutions are related to
the five dimensional solutions by considering
the case when $\tilde{c}_1 = \tilde{c}$ and/or
$\tilde{c}_2 = \tilde{c}$ in our language 
(we note our Eq. (\ref{well}) ). }.  
It will be an interesting
exercise to study the structure of the whole
solution space and the mapping between each
sector of the solution space via $U$-duality in detail.

On a broader side of our theme, the urgent next
step is to study the dynamics (including
the gravitational
back-reaction) of the supergravities via
the two dimensional dilaton gravity theories.  
Depending on whether one includes the circle
$S^1$ in the dynamical considerations, 
we end up getting (1+1)-dimensional or
(2+1)-dimensional description of the
$s$-wave dynamics.  We expect the techniques
accumulated so far in the context of the 
two dimensional dilaton gravities \cite{dilaton}
will be helpful, and the work in this regard is
in progress.

\acknowledgements{Y. K. and C.Y. L. were supported in part
by the Basic Science Research Institute Program, Ministry of
Education, BSRI-96-2442, and in part by KOSEF Grant 
971-0201-007-2. }


\appendix
\newpage
\section{Kaluza-Klein reduction on $S^1$}
In this Appendix, we set up the notation for this note and
collect some useful formulas for the Kaluza-Klein reduction
on a circle.  The signature choice for the metric 
$g^{(d+1)}_{\mu \nu} $
in $(d+1)$-dimensional
space-time in this note is $(-++...+)$, i.e., we use -1 for the 
time-like coordinate $x^0$, and +1 for a space-like coordinate 
$x^i$ (where $i = 1, ..., d$).  In $(d+1)$-dimensional 
space-time, the curvature tensor is given by
\begin{equation}
{R_{\mu \nu \rho}}^{\sigma} = 
\partial_{\nu} {\Gamma^{\sigma}}_{\mu\rho}
   -\partial_{\mu} {\Gamma^{\sigma}}_{\nu\rho}
   +{\Gamma^{\alpha}}_{\mu\rho}
   {\Gamma^{\sigma}}_{\alpha \nu}-{\Gamma^{\alpha}}_{\nu\rho}
   {\Gamma^{\sigma}}_{\alpha \mu} ,
\end{equation}
while the Christoffel connection satisfies
\begin{equation}
{\Gamma^{\rho}}_{\mu \nu}=\frac{1}{2} g^{(d+1)\rho\sigma}
   (\partial_{\mu}g^{(d+1)}_{\nu\sigma}
   +\partial_{\nu}g^{(d+1)}_{\mu\sigma}
   -\partial_{\sigma}g^{(d+1)}_{\mu\nu}) .
\end{equation}
The space-time indices $\mu$, $\nu$, $\rho$ and $\sigma$
run from zero to $d$.  For the Kaluza-Klein reduction from 
$(d+1)$-dimensions to $d$-dimensions, we write the 
metric $g^{(d+1)}_{\mu \nu}$ as
\begin{equation}
g^{(d+1)}_{\mu \nu} = \left( \begin{array}{cc}
 g^{(d)}_{\alpha \beta}+e^{\psi_1} A_{\alpha} A_{\beta}  &   
  e^{\psi_1} A_{\alpha}   \\
  e^{\psi_1} A_{\beta}    &  e^{\psi_1}   \end{array}
                      \right)  ,
\end{equation}
where the indices $\alpha$ and $\beta$ run from zero to 
$(d-1)$.  The factor $\exp ( \psi_1 /2 )$ represents the 
radius of the compact circle. The inverse of this metric is 
computed to be
\begin{equation}
g^{(d+1) \mu \nu} = \left( \begin{array}{cc}
    g^{(d) \alpha \beta}  &   - A^{\alpha}   \\
   -A^{\beta} &  e^{-\psi_1}  + g^{(d) \alpha \beta} 
A_{\alpha} A_{\beta}  \end{array}
                      \right)  ,
\end{equation}
where the $d$-dimensional indices are raised and
lowered by the $d$-dimensional metric 
$g^{(d)}_{\alpha \beta}$.  We note that the determinant
of the metric $g^{(d+1)}_{\mu \nu}$ is simply 
$\exp ( \psi_1 )$ multiplied by the determinant of
$g^{(d)}_{\alpha \beta}$.  

     For the dimensional reduction, we assume that the 
$(d+1)$-dimensional metric does not depend on the
circle coordinate $\theta \equiv x^d $.  
In other words, we only retain the
zero modes along the circle by setting
$\partial_{\theta} g^{(d+1)}_{\mu \nu} = 0 $.
By the direct computation, we verify that the 
$(d+1)$-dimensional scalar curvature $R^{(d+1)}$
decomposes into 
\begin{equation}
R^{(d+1)} = R^{(d)} - D^{(d)}_{\alpha} 
D^{(d) \alpha} \psi_1
- \frac{1}{2} D^{(d)}_{\alpha} \psi_1 
D^{(d) \alpha} \psi_1
- \frac{1}{4} e^{\psi_1} g^{(d) \alpha \beta}
g^{(d) \gamma \delta} F_{\alpha \gamma}
F_{\beta \delta} ,
\end{equation}
where $D^{(d)}_{\alpha}$ is the $d$-dimensional
covariant derivative and $F_{\alpha \beta} =
\partial_{\alpha} A_{\beta} - \partial_{\beta} 
A_{\alpha}$.  If we have a $(d+1)$-dimensional
scalar field $f$ that satisfies $\partial_{\theta} f = 0$,
we simply have
\begin{equation}
g^{(d+1) \mu \nu } D^{(d+1)}_{\mu} f 
D^{(d+1)}_{\nu} f = g^{(d) \alpha \beta}
D^{(d)}_{\alpha} f D^{(d)}_{\beta} f  .
\end{equation}
If we have a $(d+1)$-dimensional three form
field strength $H^{(d+1)}$, the kinetic term 
decomposes into
\begin{equation}
H^{(d+1) \mu \nu \rho} 
H^{(d+1)}_{\mu \nu \rho}  = 
H^{\prime (d) \alpha \beta \gamma} 
H^{\prime (d)}_{\alpha \beta \gamma}
+ 3 e^{-\psi_1} H^{(d) \alpha \beta \theta}
   H^{(d)}_{\alpha \beta \theta}
\end{equation}
where we introduce
\[ H^{\prime (d)}_{\alpha \beta \gamma} = 
H^{(d)}_{\alpha \beta \gamma}
- (A_{\alpha} H^{(d)}_{\beta \gamma \theta}
 +   A_{\beta} H^{(d)}_{\gamma \alpha \theta}
 +   A_{\gamma} H^{(d)}_{\alpha \beta \theta} ) .
\]
We can interpret $H^{(d)}_{\alpha \beta \theta}$
as a two form field strength in $d$ dimensions.

\section{Dimensional reduction with block-diagonal metric}

We consider the dimensional reduction where the metric
tensor $g^{(d)}_{\mu \nu}$ is block-diagonal.  In other
words, $g^{(d)}_{\mu \nu}$ is given by
\begin{equation}
g^{(d)}_{\mu \nu} = \left( \begin{array}{cc}
              g^{(D_l )}_{\alpha \beta}  &     0   \\
              0     &    h^{(D_t )}_{ij}   \end{array}
                      \right)  ,
\end{equation}
where the longitudinal space-time with the dimension
$d_l$ has the metric $g^{(d_l )}_{\alpha \beta}$ and the
transversal space with the dimension $d_t$ has the metric
$h^{(d_t )}_{ij}$.  Thus, we have $d = d_l + d_t$.
The indices $\alpha$ and $\beta$ run
from $0$ to $d_l -1$, whereas the indices $i$ and $j$
run from $1$ to $d_t$.  For other conventions, we follow 
the Appendix A.  This type of dimensional reduction
was utilized in Ref. \cite{reduc} for the study of 
the gravitational scatterings at Planckian energies
in four dimensional space-time.

The curvature scalar $R^{(d)}$ decomposes into 
\begin{equation}
R^{(d)} = R^{(d_l )} + R^{(d_t )}
- \frac{2}{\sqrt{- g^{(d_l)}}} h^{ (d_t) ij} D_i D_j
  \sqrt{- g^{(d_l)}}
- \frac{2}{\sqrt{h^{(d_t)}}} g^{ (d_l) \alpha \beta} 
D_{\alpha} D_{\beta}  \sqrt{h^{(d_t)}}
\label{b2}
\end{equation}
\[ - \frac{1}{4} h^{(d_t ) ij} D_i 
g^{(d_l )}_{\alpha \beta}
D_j g^{(d_l )}_{\gamma \delta} ( 
g^{(d_l ) \alpha \delta} 
g^{(d_l ) \beta \gamma} - g^{(d_l ) \alpha \beta}
g^{(d_l ) \gamma \delta} ) \]
\[ - \frac{1}{4} g^{ (d_l ) \alpha \beta} 
D_{\alpha} h^{(d_t )}_{ij}
D_{\beta} h^{(d_t )}_{kl} ( h^{(d_t) il} 
h^{(d_t )jk} - h^{(d_t ) ij}
h^{(d_t ) kl} ) \]
where $D_i$ represents the covariant derivative
in the transversal space and $D_{\alpha}$ denotes
the covariant derivative in the longitudinal 
space-time.  Here $g^{(d_l ) }$ and $h^{(d_t )}$
represent the determinants of the metric.

We consider the $d$-dimensional action
\begin{equation}
I = \frac{1}{2^{d} \pi G_d }
\int d^d x \sqrt{-g^{(d)}} ( e^{a \phi} R^{(d)}
+ 4 e^{b \phi} (D \phi )^2 -
\frac{1}{12} e^{c \phi} H^{(d) 2}  ) 
\label{b3}
\end{equation}
where we have the dilaton field $\phi$, the
three-form field strength $H^{(d)}$ and real
parameters $a$, $b$ and $c$.  In terms of the
antisymmetric two-form gauge field 
$B^{(d)}_{\mu \nu}$, we have 
$H^{(d)} = d B^{(d)} $.
We assume that the $d$-dimensional 
metric is of the form 
\begin{equation}
ds^2 = g^{(d_l )}_{\alpha \beta} 
dx^{\alpha} dx^{\beta} + e^{\psi} 
\delta^{(d_t )}_{ij} dx^i dx^j ,
\end{equation}
where $R^{(d_t )} = 0$ and
\[ \int_{(d_t )} dy^1 \cdots dy^{d_t} 
= V \]
for a constant $V$.  We further suppose
that all the relevant fields do not depend 
on the transversal coordinates.  By this assumption,
we are retaining only the zero modes of the 
Kaluza-Klein reduction.  Using Eq. (\ref{b2}),
we can integrate out the action Eq. (\ref{b3})
over the transversal space.  The result
of the integration is 
\begin{equation}
I = \frac{V}{2^d \pi G_d} 
\int d^{d_l} x \sqrt{ - g^{(d_l)} } e^{d_t \psi /2 } (
e^{a \phi} R^{(d_l)} + \frac{d_t ( d_t -1)}{4} e^{ a \phi}
(D \psi )^2 
\end{equation}
\[ + a d_t e^{a \phi} D^{\alpha} \phi
D_{\beta} \psi + 4 e^{b \phi} ( D \phi )^2
 - \frac{1}{12} e^{c \phi} H^{(d_l) 2} \]
\[ - \frac{1}{4} e^{c \phi - \psi} \delta^{(d_t) ij} 
F^{(d_l)}_i \cdot F^{(d_l)}_j
 - \frac{1}{4} e^{c \phi -2 \psi}
 \delta^{(d_t) ij} \delta^{(d_t) kl} D_{\alpha}B_{ik}
 D^{\alpha}B_{jl}) .\]
The $d$-dimensional three-form field strength 
$H^{(d)}_{\mu \nu \rho}$
decomposes into the $d_l$-dimensional three-form field
strength $H^{(d_l)}_{\alpha \beta \gamma}$, 
$d_t$ $d_l$-dimensional two-form field strength 
$F^{(d_l)}_{i \alpha \beta}$ and $d_t (d_t -1 ) /2$
one-form field strength $D_{\alpha} B_{ij}$.

\section{Derivation of general static solutions}
The differential equations Eqs.(\ref{f0}) - (\ref{s})
are quite similar to the ones considered in
Ref. \cite{kp}.  In analogy with that work, we introduce
a following set of field redefinitions
and a coordinate change
\begin{equation}
\bar{\Omega} = \exp ( \frac{3}{2 \sqrt{2}}(f_1 + f_2 ) )
\Omega \ , \
\exp ( 2 \bar{\rho} ) = \exp
( - \frac{1}{\sqrt{2}}(f_1 + f_2 ) ) \exp ( 2 \rho ) , 
\label{redef}
\end{equation}
\[ d \bar{x} = \exp ( \frac{3}{2 \sqrt{2}} (f_1 + f_2 ) )
dx , \]
which simplifies the calculations.  In terms of these
redefined fields, we can rewrite 
Eqs. (\ref{f0}) - (\ref{q}) and (\ref{s}) as follows.
\begin{equation}
f_0  = \bar{\Omega} \dot{f}  ,
\label{nf0} 
\end{equation}
\begin{equation}
\sqrt{2} f_{10}  =  \bar{\Omega} \dot{f}_a
 + QA - Q_1 A_1  \  ,  \
Q_1  \dot{A}  =  Q e^{- 2 f_a }  \dot{A}_1 , 
\label{nf10} 
\end{equation}
\begin{equation}
\sqrt{2} f_{20}  =  \bar{\Omega} \dot{f}_b 
+ QA -Q_2 A_2  \ , \
Q_2   \dot{A} =  Q e^{- 2 f_b }  \dot{A}_2 , 
\label{nf20} 
\end{equation}
\begin{equation}
Q   =  e^{-2 \bar{\rho}} 
\bar{\Omega}^{5/3}  \dot{A} ,
\label{nq} 
\end{equation}
\begin{equation}
\bar{s}   = -\frac{1}{3} \dot{\bar{\Omega}}
+ \dot{\bar{\rho}} \bar{\Omega} - Q A ,
\label{ns}
\end{equation}
where the overdot represents the differentiation
with respect to $\bar{x}$ and we introduced
$\bar{s} = s - (\sqrt{2} /  3 ) 
(f_{10} + f_{20} )$, $f_a = (2 f_1 + f_2 ) / \sqrt{2}$,
and $f_b = (2 f_2 + f_1 ) / \sqrt{2}$.  Combining
Eqs. (\ref{nq}) and (\ref{ns}), we find
\[ (2 \bar{s} + 2QA) \dot{A} =
\frac{d}{dx} ( e^{ 2 \bar{\rho} } 
\bar{\Omega}^{-2 /3} ) Q , \]
which, upon integration, becomes
\begin{equation}
Q e^{2 \bar{\rho}} \bar{\Omega}^{-2/3}
= QA^2 + 2 \bar{s} A + c \equiv P(A),
\label{rhosol}
\end{equation}
where we introduced a function $P(A)$ and
$c$ is a constant of integration.  We note that
$\epsilon$, the sign of $P(A)$, should be the
same as the sign of $Q$.   Putting
Eq. (\ref{rhosol}) into Eq. (\ref{nq}), we get
\begin{equation}
\bar{\Omega} \frac{dA}{d\bar{x}} = P(A) 
\ \ \rightarrow \ \ 
\bar{\Omega} \frac{d}{d\bar{x}} =
P(A) \frac{d}{dA} .
\label{Ax}
\end{equation}
By changing the differentiation variable from
$\bar{x}$ to $A$ with the help of Eq. (\ref{Ax}), we
immediately find that Eq. (\ref{nf0}) can be
integrated to yield
\begin{equation}
f = f_0 \int \frac{dA}{P(A)} + f_1 
\label{fsol}
\end{equation}
where $f_1$ is the constant of integration.
In a similar fashion,  we can rewrite Eq. (\ref{nf10}) as
\begin{equation}
Q \frac{dA_1}{dA} = Q_1 e^{2 f_a}
\label{aa}
\end{equation}
and 
\begin{equation}
P(A) \frac{d}{dA} f_a + QA - Q_1 A_1 =\sqrt{2} f_{10} .
\label{ab}
\end{equation}
Differentiating Eq. (\ref{ab}) with respect to
$A$ and using Eq. (\ref{aa}),
we get
\begin{equation}
 \frac{d}{dA} (P(A) \frac{d}{dA} f_a )
+ Q - \frac{Q_1^2}{Q} e^{2 f_a} = 0 .
\label{temp1}
\end{equation}
By setting 
\[ f_a = - \frac{1}{2} \ln | P(A) | + \hat{f}_a , \]
and introducing a new variable
\[ \hat{x} = \int \frac{dA}{P(A)} , \]
we find that Eq. (\ref{temp1}) is in fact 
a one-dimensional classical Liouville equation
\[ \frac{d^2}{d\hat{x}^2} \hat{f}_a - 
\frac{Q_1^2}{\epsilon Q} \exp (2 \hat{f}_a ) = 0 .
\]  
This equation can be exactly solved to determine
\begin{equation}
e^{2 f_a } = \frac{|c_1 Q|}{Q_1^2}
\frac{1}{|P(A)|} \frac{1}{\sinh^2 \left[ \sqrt{c_1}( \int 
\frac{dA}{P(A)} + \tilde{c}_1 )\right] }
\label{fasol}
\end{equation}
which, in turn, yields
\begin{equation}
Q_1 A_1 = - \sqrt{2} f_{10} - \bar{s} 
- \sqrt{c_1} \coth \left[ \sqrt{c_1} ( \int
\frac{dA}{P(A)} + \tilde{c}_1 ) \right]
\label{a1sol}
\end{equation}
upon using Eq. (\ref{ab}).  Arbitrary
real numbers $c_1$ and
$\tilde{c}_1$ are the constants of integration.
We note that if $c_1 < 0$, the hyperbolic 
functions change into the trigonometric
functions.  For $c_1 =  0$, only the leading order
term in the Taylor expansion of the hyperbolic
function survives.   

The analysis leading to Eqs. (\ref{fasol})
and (\ref{a1sol}) can be repeated for 
the fields $f_b$ and $A_2$ from Eq. (\ref{nf20}).
The results are
\begin{equation}
e^{2 f_b } = \frac{|c_2 Q|}{Q_2^2}
\frac{1}{|P(A)|} \frac{1}{\sinh^2 \left[ \sqrt{c_2} 
( \int 
\frac{dA}{P(A)} + \tilde{c}_2 ) \right] }
\label{fbsol}
\end{equation}
and
\begin{equation}
Q_2 A_2 = - \sqrt{2} f_{20} - \bar{s} 
- \sqrt{c_2} \coth \left[ \sqrt{c_2} ( \int
\frac{dA}{P(A)} + \tilde{c}_2 ) \right]
\label{a2sol}
\end{equation}
where arbitrary real number $c_2$ and
$\tilde{c}_2$ are the constants of integration.

We express Eq. (\ref{s}) in terms of the redefined
fields Eq. (\ref{redef}), change variable from $x$ 
to $\hat{x}$, plug in 
Eqs. (\ref{fasol}) - (\ref{a2sol}), to get
\begin{equation}
\frac{8}{3} ( \frac{d \hat{\phi} }{ d \hat{x} } )^2
 - \frac{1}{2 |Q|} e^{- 8 \hat{\phi} / 3} - 
\frac{3}{2} D_2 = 0
\label{ahhhh}
\end{equation}
where we introduced 
\[ 3 \bar{\Omega}^{4/3} = \exp ( - \frac{8}{3} 
\hat{\phi} -  \ln |P(A)|  ) \]
and 
\[ D_2 \equiv \frac{1}{3} ( g_0^2 + \bar{s}^2
- Qc + c_1 + c_2  ) .  \] 
The seemingly complicated Eq. (\ref{s}) drastically
simplifies after the field redefinitions.  
In fact, Eq. (\ref{ahhhh}) is the first integration of
the one-dimensional classical Liouville 
equation where $D_2$ plays the role of the
constant of the integration.  This equation
can be immediately integrated to yield
\begin{equation}
\bar{\Omega}^{4/3} = |D_2 Q|
\frac{1}{|P(A)|} 
\frac{1}{\sinh^2 \left[ \sqrt{D_2} (  
\int \frac{dA}{P(A)} + \tilde{c} )\right] }
\label{osol}
\end{equation}
where $\tilde{c}$ is a constant of integration.
Since Eq. (\ref{Ax}) determines $A$ in terms of
$x$ through
\begin{equation}
x - x_0 = \int \frac{\Omega}{P(A)} dA
\end{equation}
where $x_0$ is a constant of integration,
we have solved all the equations of motion.

\end{document}